\documentstyle{mn}

\title{On the Motion of the Local Group and its Substructures}
\author[S. Rauzy and V.G. Gurzadyan]
       {S. Rauzy$^1$ and V.G. Gurzadyan$^{2,3}$ \\
$^1$ ANPE and Centre de Physique Th\'eorique -
C.N.R.S., Luminy Case 907, F-13288 Marseille
Cedex 9, France. \\
$^2$ University of Sussex, Brighton BN1 9QH, UK.\\
$^3$ Department of 
Theoretical Physics, Yerevan Physics Institute, Yerevan
375036, Armenia (permanent address). 
}
\date{
      Received ;
      }

\begin{document}

\maketitle

 \begin{abstract}
The problem of the relative motion of the substructures of the Local
Group of galaxies revealed via S-tree method, as well as of the velocity
of the Local Group itself, is considered. The existence of
statistically significant bulk flow of 
the Milky Way  subsystem is shown via 3D reconstruction procedure, 
which uses the information on the radial velocities of the galaxies, but not 
on their distances.
Once the bulk motion of substructures is estimated, in combination with the 
observed CMB dipole we also consider the mean velocity of the Local Group 
itself. 
Assigning the Local Group the mean motion of its main substructures we evaluate
its peculiar velocity in Milky Way frame
${\rm V}_{{\rm LG} \rightarrow {\rm MW}}=
(-7 \pm 303,-15 \pm 155 ,+177  \pm 144)$
or $178$ km s$^{-1}$ toward galactic coordinates $l=245$ and $b=+85$.
Combined with CMB dipole 
${\rm V}_{{\rm MW} \rightarrow {\rm CMB}}$, we obtain Local Group
velocity in CMB frame:
${\rm V}_{{\rm LG} \rightarrow {\rm CMB}}$ = 
($-41\pm303,-497\pm 155,445 \pm 144)$ or  
$668$ km s$^{-1}$ towards $l=265$ and $b=42$.
This estimation is in good agreement, within 1 $\sigma$ level, with the 
estimation of Yahil et al (1977).
 \end{abstract}

\section{Introduction}
\label{Introduction}

The peculiar motion of the Local Group (LG) seems to become one of the
essential problems of observational cosmology, since it can provide
at least crucial contraints on properties of the local region
of the Universe. In more optimistic evaluation, it can be the cornerstone
for the linear theory of gravitational instabilities and, hence, for
many aspects of the Big Bang cosmology, in general. 

The important question here is, obviously, the convergence of various dipoles,
e.g. of optical galaxies and
clusters of galaxies, IRAS galaxies, X-ray clusters, X-ray galaxies, etc.,
with the dipole of the Cosmic Microwave Background radiation, if the latter
is caused by the Doppler effect.
The reliable determination of the LG CMB dipole itself is 
the central point here.

Though there is no absolute convergence, all the dipoles, in general,
either are aligned in the direction of the defined CMB dipole or differ within
limits, which presumably are affected by the choice of
particular samples. For example, it is noticed, that the convergence level 
is improved, if deeper samples of extragalactic objects are involved, 
i.e. the matter distribution on more larger cosmic volumes is
taken into account (Branchini\&Plionis (1996)).

While the result by Lauer\&Postman (1994) on the quite differently
oriented dipole obtained from the clusters of galaxies with 
velocities up to 15 000 km s$^{-1}$, seems to challenge the situation,
the subsequent analysis by Branchini\&Plionis (1996) involving even deeper
survey of Abell/ACO clusters - up to 20 000 km s$^{-1}$, showed alignment
within  $10^{\circ}$ with respect to the CBM dipole, if the Virgocentric
flow is also taken into account. 

The main common concern in these results based on the analysis of samples 
of various types of extragalactic objects, is in which degree the given
population really traces the large-scale mass distribution. Though the
general alignment of the dipoles of a given population with the CMB one,
can imply the positive answer to this question, the reliably obtained
divergence level of any of dipoles can be not less informative.     
Therefore, any alternative means of estimation of the motion of the LG
should be of particular interest.

In the present study we consider the possibility to estimate 
the peculiar velocity of the LG using not the extragalactic
information, but its internal dynamical properties combined with data on
CMB dipole. Namely, we analyse
the substructure of the LG, and perform a procedure of 3D
reconstruction  of the velocity of the whole system and its main subgroups.
Thus, we continue to use the approach of 3D reconstruction of the
tangential velocities of galaxy systems, developed in Gurzadyan\&Rauzy
(1997), with the difference that in the present study the line-of-sight
velocities have no components representing the Hubble flow.

The {\it first} step is performed by the S-tree technique 
(see Gurzadyan\&Kocharyan (1994)) enabling,
in particular, the analysis of systems 
such as groups or clusters of galaxies. 
For 
the {\it second} step, we use the data on the distribution of the line-of-sight
velocities of the galaxies which are members of the LG and its
subsystems, as revealed by the first step.
Thus, we obtain the peculiar motions  of the Milky
Way and M31 subsystems with respect to each other.
By the {\it third} step,
from the bulk motions of the mentioned
both subsystems we obtain the velocity vector of the
Local Group in CMB frame. 
These results are discussed in term of the convergence
of the various luminous dipoles with the one observed
in the Cosmic Microwave Background Radiation.

\section{The S-tree technique}

The S-tree technique is developed for the investigation of properties of 
many-dimensional nonlinear systems and essentially uses the concepts 
of the theory of dynamical systems. Refering for details to 
(Gurzadyan et {\it al} 1991, 1994; Gurzadyan\&Kocharyan 1994
Bekarian\&Melkonian 1997),
 here we outline its key points only. 
Its idea is based on the property of structural stability
well known in theory of dynamical systems enabling to study 
the robust properties of the systems with limited amount of information.
The gravitating systems were known to be
exponentially unstable and hence being among systems with strong statistical
properties.
The advantage of the method in the context of galaxy clusters is in the 
self-consistent  use of both kinematical and positional information
on the clusters, as well as of the data on individual observable
properties of galaxies - the magnitudes.

This approach is introducing the concept
of the {\it degree of boundness} between the members of the given 
N-body system, i.e. the definition of a nonnegative 
function ${\cal P}$ called the boundness function
which describes the degree of interaction of a given subset $Y$ of the
initial set $\cal A$ with another subset $X\backslash Y$ 
according to some criterion.
Particularly one can define a function
$$
    {\cal P}:{\cal S(A)}\to R_+:(X,Y)\mapsto {\cal P}_X(Y),
$$
where $\cal S(A)$ is the set of all subsets of the initial set, 
when one of the sets contains the other one. 
The procedure of such splitting of ${\cal A}$ can be measured by a 
non-negative real number $\rho$ using the boundness function,
so that the problem is reduced to the finding out of a
function $\Sigma(\rho)$ denoting the set of all possible 
$\rho$-subsystems
$\{{\cal A}_1,\ldots,{\cal A}_d\}$ of $({\cal A, P})$. 

     Attributing for the given $\rho$ the matrix $D$ to 
another matrix $\Gamma$  in a following way:
\begin{equation}
\begin{array}{cll}
\Gamma_{ab}=0 \mbox{ if } &D_{ab} <    \rho \mbox{, } &D_{ba} < \rho,\\
\Gamma_{ab}=1 \mbox{ if } &D_{ab} \geq \rho \mbox{, } &D_{ba} \geq \rho;
\end{array}
\end{equation}
the problem of the search of a $\rho$-bound cluster is reduced  to  that 
of the connected parts  of  the  graph $\Gamma(\rho)$ - a 
{\em S-tree} diagram.

For the matrix $D_{ab}$ a representation via the two-dimensional
curvature $K$ of the phase space of the system can be used: 
$$
    D_{ab}=\max_{i,j}\left\{-K^{\mu}_{ \nu},0\right\},
$$
where $\mu=(a,i), \nu=(b,j)$. The two-dimensional curvature is represented
by the Riemann tensor $R$ via the expression: $K_{\mu\nu}=R_{\mu\nu\lambda\rho}
u^{\lambda}u^{\rho}$, where $u^{\nu}$ is the velocity of the geodesics; the
explicit expression of two-dimensional curvature for N-body gravitating
system is derived in (Gurzadyan \&Savvidy, 1986) and has rather
complicated form to be represented here.
The advantage of the use of geometric characteristics such as Riemann, Ricci 
tensors, is well known in theory in dynamical systems, and has been used
in  astrophysical problems as well.

As a result, the degree of boundness between the members and subgroups
of the given N-body system can be obtained, thus revealing the physically 
interacting system (cluster) and its hierarchical substructure.  
 
The computer code based on the S-tree method  has been  used for the 
study of the substructure of the Local Group of galaxies
(Gurzadyan et al. (1993)), of the core
of Virgo (Petrosian et al. 1997) and Abell clusters
(Gurzadyan \& Mazure, 1997).
In these studies the information on the masses of galaxies has been
used via the mass-to-luminosity ratio $M={\rm const}\, L^n$, $n=1$, though other
relations -- ($n=0, 1/2$), have been checked as well, and the robust character
of the results on the subgrouping has been revealed; note, that
these assumptions take into account also the existence
of the dark matter associated with the galaxies.

\section{Bulk flow reconstruction}
\label{bulkflow}

The reconstruction of the 3D velocity distribution function
from its observed line-of-sight velocity distribution  
is an interesting inverse problem.
It was analytically solved by 
Ambartsumian (1936) for stellar systems without any {\it a priori} assumption
on the form of the soght function.
The only assumption made
was the independence of the distribution function on   
the spatial regions (directions), or equivalently, that 
the 3D velocity distribution 
$\phi(v_1,v_2,v_3)~dv_1\,dv_2\,dv_3$ of such  
systems was invariant under spatial translations.
It means, that the theoretical probability density $dP_{th}$
of the system reads as follows 
\begin{equation}
\label{dpth}
dP_{\rm th}=\phi(v_1,v_2,v_3)\,dv_1\,dv_2\,dv_3
~\rho(r,l,b)\,
r^2\cos b \,dr\,dl \,db
\end{equation}
where 
$\phi(v_1,v_2,v_3)$ is the 3D velocity distribution function (in
galactic cartesian coordinates) and $\rho(r,l,b)$ is the 3D spatial
distribution function (in galactic coordinates).
Under this assumption, Ambartsumian had proven the 
theoretical possibility of reconstructing of 
$\phi(v_1,v_2,v_3)$ from the observed radial 
velocity distribution function
$f(v_r,l,b)$ 
\begin{equation}
\label{vrlbdistribution}
dP_{\rm obs} = f(v_r,l,b)~\cos b\, dl\, db\, dv_r
\end{equation}
where
$v_r={\bf v.{\hat r}}$ is the projection of the velocity
${\bf v}=(v_1,v_2,v_3)$ on the line-of-sight direction
${\bf {\hat r}}=
({\hat r}_1,{\hat r}_2,{\hat r}_3)=
(\cos l\cos b,\sin l\cos b,\sin b)$.

However, computer experiments show that the direct application of the 
Ambartsumian's formula is hardly possible for that purpose.
It is natural, since the
derivation of a smooth function based on  discrete
information on relatively small number of particles (say, less than 1000)
is a nonlinear problem. The number of points should exceed essentially
the number of galaxies of real clusters
in order to apply this formula with success.
This fact is a consequence of  the principal difference
between the N-body problems in stellar dynamics and dynamics of clusters
of galaxies. 

Nevertheless, some quantities of interest such as the first
order moment of this distribution, i.e. the mean
3D velocity of the system, can be obtained.
Additional hypotheses on the distribution function
$\phi(v_1,v_2,v_3)$ are then necessary. 
Hereafter, we assume that 
$\phi(v_1,v_2,v_3)$ can be written as follows 
\begin{equation}
\label{phi3D}
\phi(v_1,v_2,v_3)~dv_1\,dv_2\,dv_3=
\prod^3_{i=1} \phi_i(v_i)\,dv_i
\end{equation}
where the distribution functions $\phi_i$ are centered on 
${\overline v}_i$ and of variances $\sigma_i$
($\phi_i(v_i)=\phi_i(v_i;{\overline v}_i,\sigma_i)$).
The statistics of the mean 3D  velocity, i.e. the bulk flow velocity,
are derived in the appendix A.
The use of the maximum liklehood technique forces us to
entirely specify  the functions $\phi_i$. We choose
the Gaussian representation ($\phi_1=\phi_2=\phi_3=g$ with $g$ Gaussian) and
assume the isotropic velocity dipersions
($\sigma_1=\sigma_2=\sigma_3=\sigma_v$).
The velocity field  of the system is, thus,
split into a mean 3D velocity
${\bf v}_B=(v_x,v_y,v_z)=({\overline v}_1,{\overline v}_2,{\overline v}_3$)
(i.e. a bulk flow) plus a 3D random
component, isotropic Maxwellian, centered on ${\bf 0}$ and
of velocity dispersion $\sigma_v$. 

The fact that gravitating N-body systems do posess strong statistical
properties peculiar to Kolmogorov systems (Gurzadyan\&Savvidy, 1986; see
also Gurzadyan\&Pfenniger, 1994),
can be the justification for the separable form of
distribution functions in Eqs.(2) and (4).
Note, that the representation (4) is valid if the potential is time independent,
and the energy is an isolating integral. 
The exact Maxwellian form of the distribution, however, is not required for 
the analysis, its representation in the form $g(v_i,{\overline v}_i,\sigma)$
is sufficient.

The derived bulk flow statistics is a robust estimator.
It does not depend on the distances of galaxies under 
consideration. This is a positive feature, since there is 
still disagreement between various authors in the distance
estimates of the members of LG.
Moreover, it is shown in appendix A
that the 
${\tilde v_x}$,
${\tilde v_y}$ and 
${\tilde v_z}$ estimators do not depend on
the projected angular
distribution function $\eta(l,b)$ of the sample,
even if their accuracies do. 

\section{Application to the Local Group}
\label{localgroup}

\subsection{The Local Group subsystems}
\label{lgsubsystems}

\begin{table}
\label{lgdata}
\begin{center}
\caption[]{
Local Group data used for this analysis:  the equatorial
coordinates, radial velocity $v_r$
(WITH RESPECT TO THE MILKY WAY REST FRAME) and their distance r, are listed
for 32 objects within 2 Mpc. M31 and Milky Way subsystems
were deduced by using the S-tree  method 
(Gurzadyan et {\it al.} (1993). Some typographical mistakes
occured in Table 1 of the mentioned paper are
corrected in the present Table).
}
\begin{tabular}{|lrrrl|}
\hline
\multicolumn{5}{|c|}{}  \\ 
\multicolumn{5}{|c|}{{\small {\bf LOCAL GROUP :}} 32 galaxies}  \\ 
\multicolumn{5}{|c|}{}  \\ 
\hline
\multicolumn{5}{|c|}{}  \\ 
~~~Object&
$\alpha$~~~~~&
$\delta$~~~~~&
$v_r$~~~~~
&
~~~r \\
&
(1950)~&
(1950)~&
(km s$^{-1}$)&
(Mpc)\\
\hline
\multicolumn{5}{|c|}{}  \\ 
\multicolumn{5}{|c|}{Milky Way subsystem : 12 galaxies}  \\ 
\multicolumn{5}{|c|}{}  \\ 
\hline
Milky Way     &00\ 00.00    & 00\ 00.0     & 0~~~~~~           &0.0\\
SMC           &00\ 51.00    &$-$73\ 06.0     &$-$21$\pm$5~~     &0.063\\
Sculptor S    &00\ 57.60    &$-$33\ 58.0     & 74$\pm$?~~         &0.085\\
IC1613        &01\ 02.22    & 01\ 51.0     &$-$153$\pm$3~~     &0.64\\
Fornax D      &02\ 37.84    &$-$34\ 44.4     &$-$51$\pm$?~~          &0.19\\
LMC           &05\ 24.00    &$-$69\ 48.0     &130$\pm$?~~          &0.052\\
Carina        &06\ 40.40    &$-$50\ 55.0     &14$\pm$?~~          &0.093\\
Leo I         &10\ 05.77    & 12\ 33.2     &177$\pm$?~~          &0.22\\
Leo II        &11\ 10.83    & 22\ 26.1     &15$\pm$?~~           &0.22\\
UM            &15\ 08.20    & 67\ 18.0     &$-$88$\pm$?~~         &0.065\\
Draco         &17\ 19.40    & 57\ 58.0     &$-$95$\pm$?~~         &0.075\\
N6822         &19\ 42.12    &$-$14\ 55.7     &55$\pm$5~~       &0.62\\
\hline
\multicolumn{5}{|c|}{}  \\ 
\multicolumn{5}{|c|}{M31 subsystem : 7 galaxies}  \\ 
\multicolumn{5}{|c|}{}  \\ 
\hline
N147          &00\ 30.46    & 48\ 13.8     & 36$\pm$50~    &0.7\\
N185          &00\ 36.19    & 48\ 03.7     & 67$\pm$22~     &0.72\\
N205          &00\ 37.64    & 41\ 24.9     & 49$\pm$11~     &0.64\\
M31           &00\ 40.00    & 40\ 59.7     &$-$110$\pm$1~~      &0.67\\
M32           &00\ 39.97    & 40\ 35.5     & 86$\pm$6~~      &0.66\\
LGS-3         &01\ 01.20    & 21\ 37.0     &$-$100$\pm$8~~     &0.72\\
M33           &01\ 31.05    & 30\ 23.9     &$-$39$\pm$0.5    &0.82\\
\hline
\multicolumn{5}{|c|}{}  \\ 
\multicolumn{5}{|c|}{Others : 13 galaxies (with r $\le$ 2 Mpc)}  \\ 
\multicolumn{5}{|c|}{}  \\ 
\hline
IC10          &00\ 17.69    & 59\ 00.9     &$-$144$\pm$2~~      &1.25\\
N55           &00\ 12.40    &$-$39\ 28.0     & 97$\pm$3~~      &1.4\\
IC342         &03\ 41.95    & 67\ 56.4     &221$\pm$?~~          &1.84\\
Leo A         &09\ 56.53    & 30\ 59.2     &$-$17$\pm$10~     &1.59\\
N3109         &10\ 00.80    &$-$25\ 54.8     &193$\pm$?~~          &1.6\\
Sex A         &10\ 08.57    &$-$04\ 27.7     &154$\pm$5~~      &1.3\\
DDO155        &12\ 56.20    & 14\ 29.2     &183$\pm$5~~      &1.0\\
Sagittarius   &19\ 27.01    &$-$17\ 47.0     &10.9$\pm$1.4   &1.11\\
DDO210        &20\ 44.13    &$-$13\ 02.0     &$-$13$\pm$10~     &1.5\\
IC5152        &21\ 59.60    &$-$51\ 32.0     &78$\pm$15~      &1.51\\
Hog           &23\ 23.80    &$-$32\ 40.0     &75$\pm$5~~       &1.3\\
Pegasus       &23\ 26.05    & 14\ 28.3     &$-$14$\pm$7~~      &1.0\\
W-L-M         &23\ 59.40    &$-$15\ 44.6     &$-$56$\pm$8~~       &1.0\\
\hline
\end{tabular}
\end{center}
\end{table}

In the S-tree study of the substructure of the LG 
(Gurzadyan et {\it al.} (1993)) a
sample of 39 galaxies has been used, compiled from various studies.
The membership of galaxies in two main subgroups  -
dominated by the Milky Way and of M31, has been revealed, in general, 
confirming the conventional views. Physical connections between 
some individual galaxies have been also indicated, not reported before 
(e.g. of NGC 6822 and IC 1613). 
Some galaxies appeared to have no actual influence on the dynamics of 
LG and vice versa, and therefore, were considered not to be its 
members. 
The LG as a physical system of the influenced galaxies only, was 
shown to extend on less than 2 Mpc. 
These galaxies and their membership are listed Table 1.  
The recent studies (see Karachentsev (1995) and references therein), though
record some differences in the data on the galaxies, in general, no
drastic reliable change is observed. 
Existence of galaxies obscured by the Galactic disk, like newly 
discovered Dwingeloo 1,2 ones
(Burton et {\it al.}  (1996)), also cannot be ruled out, 
though it seems unlikely that, at least the latter ones can essentially 
influence the substructure of the LG (Lynden-Bell, 1996).

The choice of membership criteria is always crucial while studying the 
substructures of clusters of galaxies.  
In Karachentsev (1995), 
only closed Keplerian orbits are considered as criteria of a membership,
while strictly speaking, precise Keplerian closed orbits are never typical
in few-body problem, as one has in the case of LG.
As it is mentioned above, the S-tree technique enables to reveal the
degree of influence, whatever the exact shape of the orbits is.

Our aim here is to show the principal possibility of the reconstruction
of 3D vector of the LG based on the substructure analysis, 
having in mind, that the latter can be easily revised, if some observational
input data on galaxies should be essentially modified with high enough
confidence level.   

\begin{table}
\label{MWresult}
\begin{center}
\caption[]{
Estimates of the bulk flow and of the velocity
dispersion for the Milky Way subsystem. Residuals
are listed for each object. Standard deviations 
are computed by using numerical simulations. 
}
\begin{tabular}{|lrr|}
\hline
\multicolumn{3}{|c|}{Milky Way subsystem : 12 galaxies}  \\ 
\multicolumn{3}{|c|}{}  \\ 
\hline
\multicolumn{3}{|c|}{{\small {\bf Bulk flow :}} 
~~${\bf v}_B=(v_x,v_y,v_z)$}  \\ 
\multicolumn{3}{|c|}{}  \\ 
\hline
\multicolumn{2}{|c|}{Galactic coordinates}&
\multicolumn{1}{|c|}{Equatorial}\\
\multicolumn{1}{|c}{(cartesian)}&
\multicolumn{1}{c|}{}&
\multicolumn{1}{|c|}{(1950)}\\
\hline
$v_x=~~104\pm64$ km s$^{-1}$ &
$l=312$~~~ & $\alpha=~~13~ 38.75$~ \\
$v_y=-116\pm42$ km s$^{-1}$ &
$b=16\,$~~~~ & $\delta=-45~43.8$~~ \\
\multicolumn{1}{|l}{$v_z=~~~45\pm33$ km s$^{-1}$} & 
\multicolumn{2}{c|}{$\|{\bf v}_B\|=162$ km s$^{-1}$~~~}\\ 
\hline
\multicolumn{3}{|c|}{{\small {\bf Velocity dispersion :}} 
~~$\sigma_v=69\pm15$ km s$^{-1}$}\\ 
\multicolumn{3}{|c|}{}  \\ 
\hline
\multicolumn{3}{|c|}{}  \\ 
\multicolumn{3}{|c|}{{\small {\bf Residuals}}}  \\ 
~Object & $v_r$~~~~~~ &$v_r-{\bf{\hat r}.v}_B$~~ \\
\hline 
\multicolumn{3}{|c|}{$\sigma_v^{\star}$
=58 km s$^{-1}$} \\ 
\hline 
~SMC           &$-21$~~~~~ &$-99$~~~~~~~ \\
~Sculptor S    &$74$~~~~~ &$107$~~~~~~~ \\
~IC1613        &$-153$~~~~~ &$-37$~~~~~~~ \\
~Fornax D      &$-51$~~~~~ &$-23$~~~~~~~ \\
~LMC           &$130$~~~~~ &$45$~~~~~~~ \\
~Carina        &$14$~~~~~ &$-55$~~~~~~~ \\
~Leo I         &$177$~~~~~ &$76$~~~~~~~ \\
~Leo II        &$15$~~~~~ &$-25$~~~~~~~ \\
~UM            &$-88$~~~~~ &$-21$~~~~~~~ \\
~Draco         &$-95$~~~~~ &$-31$~~~~~~~ \\
~N6822         &$55$~~~~~ &$31$~~~~~~~ \\
\hline 
\end{tabular}
\end{center}
\end{table}
\begin{table}
\label{M31result}
\begin{center}
\caption[]{
Estimates of the bulk flow and of the velocity
dispersion for the M31 subsystem. Residuals
are listed for each object. Standard deviations 
are computed by using numerical simulations. 
}
\begin{tabular}{|lrr|}
\hline
\multicolumn{3}{|c|}{M31 subsystem : 7 galaxies}  \\ 
\multicolumn{3}{|c|}{}  \\ 
\hline
\multicolumn{3}{|c|}{{\small {\bf Bulk flow :}} ~~
${\bf v}_B=(v_x,v_y,v_z)$}  \\ 
\multicolumn{3}{|c|}{}  \\ 
\hline
\multicolumn{2}{|c|}{Galactic coordinates}&
\multicolumn{1}{|c|}{Equatorial}\\
\multicolumn{1}{|c}{(cartesian)}&
\multicolumn{1}{c|}{}&
\multicolumn{1}{|c|}{(1950)}\\
\hline
$v_x=-117\pm541$ km s$^{-1}$ &
$l=144$~~~ & $\alpha=11~ 54.35$~ \\
$v_y=~~~85\pm267$ km s$^{-1}$ &
$b=65\,$~~~~ & $\delta=50~07.09$~ \\
\multicolumn{1}{|l}{$v_z=~~309\pm254$ km s$^{-1}$} & 
\multicolumn{2}{c|}{$\|{\bf v}_B\|=341$ km s$^{-1}$~~~}\\ 
\hline
\multicolumn{3}{|c|}{{\small {\bf Velocity dispersion :}} 
~~$\sigma_v=78\pm20$ km s$^{-1}$} \\ 
\multicolumn{3}{|c|}{}  \\ 
\hline
\multicolumn{3}{|c|}{}  \\ 
\multicolumn{3}{|c|}{{\small {\bf Residuals}}}  \\ 
~Object & $v_r$~~~~~~ &$v_r-{\bf{\hat r}.v}_B$~~ \\
\hline 
\multicolumn{3}{|c|}{$\sigma_v^{\star}$
=57 km s$^{-1}$} \\ 
\hline 
~N147          &$36$~~~~~ &$-16$~~~~~~~ \\
~N185          &$67$~~~~~ &$15$~~~~~~~ \\
~N205          &$49$~~~~~ &$36$~~~~~~~ \\
~M31           &$-110$~~~~~ &$-120$~~~~~~~ \\
~M32           &$86$~~~~~ &$78$~~~~~~~ \\
~LGS-3         &$-100$~~~~~ &$-2$~~~~~~~ \\
~M33           &$-39$~~~~~ &$0$~~~~~~~ \\
\hline 
\end{tabular}
\end{center}
\end{table}
\subsection{Bulk flow estimate}
\label{lgbulkflows}

The MW and M31 subsystems contain respectively 12 and 7
galaxies. Such a small number of  objects forbids
indeed the use of the Ambarsumian's formulae
in order to reconstruct the 3D velocity distribution
functions of these two systems.
Neverthless, in this section we show that it is possible
to evaluate their 3D mean velocity, i.e. their bulk flow.

For each subsystem, as it is assumed Eq. (2),
the velocity distribution function $\phi(v_1,v_2,v_3)$
is translation invariant, so that 
$\phi(v_1,v_2,v_3)$ can be split into a bulk flow
${\bf v}_B=(v_x,v_y,v_z)$ plus a 3D random
component, isotropic Maxwellian, centered on ${\bf 0}$ and
of velocity dispersion $\sigma_v$ (see section \ref{bulkflow}, Eq.(4)). 
According to the bulk flow statistics derived in appendix A,
Tables 2 and 3 give
the values of the bulk flow estimates for the MW and M31 subsystems,
respectively.
These velocities are expressed in km s$^{-1}$ and with
respect to rest frame of the Milky Way (MWR frame).

For each system, the values of the velocity
dipersion $\sigma_v$ and of the
accuracy of the bulk flow estimate are calculated
by using numerical simulations. Large number of samples have been
simulated according to the following characteristics:
\\\indent - The angular position of
each object of the simulated system is identical with the observed
one. 
\\\indent - The 3D velocity of each object is the
estimated bulk flow (for example, ${\bf v}_B=(104,-116,45)$
in km s$^{-1}$
for the MW subsystem) plus a 3D random component, Maxwellian
and isotropic, of velocity dispersion $\sigma_v$.
\\\indent - The simulated
radial velocity of each object is the line-of-sight projection
of its 3D velocity, plus a white noise accounting for
measurment errors on $v_r$ (see table 1). When this
information is missing, a value of 10 km s$^{-1}$ is adopted.

Since the quantity $\sigma_v$ is unknown, we adopt
the value of the biased estimator $\sigma_v^\star$ of $\sigma_v$
as a fiducial starting point (see appendix A). 
After few iterations, the correct value of
the velocity dispersion $\sigma_v$ is reached when
the $\sigma_v^\star$
estimate of the simulated samples corresponds to the observed one.
Errors bars on the $v_x$, $v_y$, $v_z$ and $\sigma_v$ estimates
are afterwards
calculated by using a large number of simulated samples.

\begin{table}
\label{LGresult}
\begin{center}
\caption[]{
Estimates of the bulk flow and of the velocity
dispersion for the Local Group, ignoring its
substructures. Residuals
are listed for each object and the $\sigma_v^{\star}$
estimates are given for the M31 and Milky Way subsystems and for
the 13 remaining galaxies. Standard deviations 
are computed by using numerical simulations. 
}
\begin{tabular}{|lrr|}
\hline
\multicolumn{3}{|c|}{LOCAL GROUP : 31 galaxies (with r $\le$ 2 Mpc)}  \\ 
\multicolumn{3}{|c|}{}  \\ 
\hline
\multicolumn{3}{|c|}{{\small {\bf Bulk flow :}} 
~~${\bf v}_B=(v_x,v_y,v_z)$}  \\ 
\multicolumn{3}{|c|}{}  \\ 
\hline
\multicolumn{2}{|c|}{Galactic coordinates}&
\multicolumn{1}{|c|}{Equatorial}\\
\multicolumn{1}{|c}{(cartesian)}&
\multicolumn{1}{c|}{}&
\multicolumn{1}{|c|}{(1950)}\\
\hline
$v_x=-28\pm45$ km s$^{-1}$ &
$l=233$~~~ & $\alpha=08~ 13.35$~ \\
$v_y=-37\pm32$ km s$^{-1}$ &
$b=13$~~~~~ & $\delta=-11~00.53$~ \\
\multicolumn{1}{|l}{$v_z=-11\pm30$ km s$^{-1}$} & 
\multicolumn{2}{c|}{$\|{\bf v}_B\|=47$ km s$^{-1}$~~~~}\\ 
\hline
\multicolumn{3}{|c|}{{\small {\bf Velocity dispersion :}} 
~~$\sigma_v=103\pm14$ km s$^{-1}$} \\ 
\multicolumn{3}{|c|}{}  \\ 
\hline
\multicolumn{3}{|c|}{{\small {\bf Residuals}}}  \\ 
~Object & $v_r$~~~~~~ &$v_r-{\bf{\hat r}.v}_B$~~ \\
\hline 
\multicolumn{3}{|c|}{Milky Way subsystem : $\sigma_v^{\star}
=73$ km s$^{-1}$} \\ 
\hline 
~SMC           &$-21$~~~~~ &$-25$~~~~~~~ \\
~Sculptor S    &$74$~~~~~ &$86$~~~~~~~ \\
~IC1613        &$-153$~~~~~ &$-138$~~~~~~~ \\
~Fornax D      &$-51$~~~~~ &$-60$~~~~~~~ \\
~LMC           &$130$~~~~~ &$109$~~~~~~~ \\
~Carina        &$14$~~~~~ &$-21$~~~~~~~ \\
~Leo I         &$177$~~~~~ &$78$~~~~~~~ \\
~Leo II        &$15$~~~~~ &$13$~~~~~~~ \\
~UM            &$-88$~~~~~ &$-76$~~~~~~~ \\
~Draco         &$-95$~~~~~ &$-69$~~~~~~~ \\
~N6822         &$55$~~~~~ &$98$~~~~~~~ \\
\hline 
\multicolumn{3}{|c|}{M31 subsystem : $\sigma_v^{\star}
=78$ km s$^{-1}$} \\ 
\hline 
~N147          &$36$~~~~~ &$56$~~~~~~~ \\
~N185          &$67$~~~~~ &$87$~~~~~~~ \\
~N205          &$49$~~~~~ &$69$~~~~~~~ \\
~M31           &$-110$~~~~~ &$-90$~~~~~~~ \\
~M32           &$86$~~~~~ &$106$~~~~~~~ \\
~LGS-3         &$-100$~~~~~ &$-83$~~~~~~~ \\
~M33           &$-39$~~~~~ &$-27$~~~~~~~ \\
\hline 
\multicolumn{3}{|c|}{Others (with r $\le$ 2 Mpc) : $\sigma_v^{\star}
=121$ km s$^{-1}$} \\ 
\hline 
~IC10           &$144$~~~~~ &$-163$~~~~~~~ \\
~N55          &$97$~~~~~ &$110$~~~~~~~ \\
~IC342         &$221$~~~~~ &$222$~~~~~~~ \\
~Leo A         &$-17$~~~~~ &$-49$~~~~~~~ \\
~N3109         &$193$~~~~~ &$151$~~~~~~~ \\
~Sex A         &$154$~~~~~ &$113$~~~~~~~ \\
~DDO155        &$183$~~~~~ &$170$~~~~~~~ \\
~Sagittarius   &$10.9$~~~~~ &$52$~~~~~~~ \\
~DDO210        &$-13$~~~~~ &$30$~~~~~~~ \\
~IC5152        &$78$~~~~~ &$98$~~~~~~~ \\
~Hog           &$75$~~~~~ &$97$~~~~~~~ \\
~Pegasus       &$-14$~~~~~ &$19$~~~~~~~ \\
~W-L-M         &$-56$~~~~~ &$97$~~~~~~~ \\
\hline 
\end{tabular}
\end{center}
\end{table}

The effect of the Hubble expansion have been also investigated.
The radial velocities listed in table 1 are indeed not corrected
for the Hubble flow. We have applied this correction by substracting 
to the radial velocity $v_r$ of each objects its
Hubble velocity $H_0r$. Since the LG galaxies
are close to the Milky Way, the bulk flow estimate
remains almost unchanged. For $H_0=100$ km s$^{-1}$ Mpc$^{-1}$,
the MW bulk flow becomes ${\bf v}_B=(106,-128,60)$ 
(i.e. $v_B=176$ km s$^{-1}$ toward $l=310$ and $b=20$) and
the M31 bulk flow 
${\bf v}_B=(-2,77,327)$
(i.e. $v_B=336$ km s$^{-1}$ toward $l=91$ and $b=77$).

In Table 3, bulk flow components and associated
uncertainties of the M31 subsystem
are expressed in galactic cartesian
coordinates. In the coordinates frame such that
X-axis is aligned with M31 line-of-sight (i.e.
toward $l=121.2$ and $b=-21.6$) and Y-axis lies in the
galactic plane, the 3D bulk flow rewrites
${\bf v}_B=(10 \pm 34,72 \pm 623 ,335  \pm291 )$.
In this coordinates, bulk flow of the MW subsystem reads
${\bf v}_B=(-155 \pm 57,-26 \pm 53 ,-15  \pm 26 )$.

In Table 2, one can notice 
that 3D bulk flow estimates of the MW subsystem does not take
into account the radial velocity of Milky Way itself.
Since the analysis is performed in the MW rest frame,
the MW velocity is zero by definition.
However, no prefered
line-of-sight direction can be assigned to the Milky Way
by a MW observer. In order to overcome this principal
difficulty, let us proceed to the following thought experiment.
The observer ventures a footstep outside Milky Way, small
enough such that the radial velocities of others galaxies remain
unchanged. It allows indeed to define
a line-of-sight
direction for  MW, opposite to the footstep walk.
The bulk flow statistic derived appendix A is then applied 
to the MW subsystem, including the zero radial velocity 
of MW with its associated direction.
This scheme is repeated for a large number
of random footstep directions, such that the MW line-of-sight
direction distribution becomes isotropic.
Averaging over all these bulk flow estimates finally gives 
the 3D mean velocity of MW subsystem.
The results of Table 2 remain almost unchanged, i.e. 
${\bf v}_B=(81\pm 68, -104 \pm 44, 38 \pm 34)$
or $v_B=137$ km s$^{-1}$ pointing toward $l=308$ and $b=16$,
with a velocity dispersion of
$71 \pm 16$ km s$^{-1}$.

We have also performed our analysis on the 32 Local
Group galaxies situated nearer than 2 Mpc, ignoring the
MW and M31 dynamical substructures. Bulk flow and
velocity dispersion estimates are shown in Table 4.
Note, however, that these estimates have no much sense since
the presence of the MW and M31 subsystems rules
out our main assumption, i.e. the velocity distribution
of the 32 Local Group galaxies is not invariant under
spatial translations. 
The values of the biased velocity dispersion estimator
$\sigma_v^\star$ for the MW and M31 subsamples are
more interesting. They are significantely greater
than the
$\sigma_v^\star$ estimates of tables 2 and 3.
This fact indicates
the existence of kinematic substructures in the Local Group, and so, 
strengthens the present analysis.

\section{Conclusion and discussion}

The motions of the two subsystems of the Local Group
have been estimated. The M31 and MW dynamical subsystems,
containing respectively 7 and 12 galaxies, have been identified
via the S-tree technique, which takes into account in
a self-consistent way the degree of influence of one
object (or a set of objects) on another, whatever their orbits can be.
In the rest frame of the Milky Way (MWR frame) the bulk flow
statistic derived appendix A, gives
${\bf v}_B=(104 \pm 64,-116 \pm 42, 45 \pm 33)$
in km s$^{-1}$ and in cartesian galactic coordinates
(or $v_B=162$ km s$^{-1}$ pointing toward $l=312$ and $b=16$)
for the motion of the MW subsystem, and
${\bf v}_B=(-117 \pm 541,85 \pm 267 ,309  \pm254 )$
or $v_B=341$ km s$^{-1}$ toward $l=144$ and $b=65$
for the motion of the subsystem dynamically associated
with M31.
While these estimates have been derived from the radial
velocity and angular position of galaxies,
the information on the distance has not been used
(the point has its importance since the main ambiguity
in the data concerns the galaxy distances). Note, that
the independence on the distances concerns only the velocity
reconstruction procedure, while the information on distances
is used for S-tree analysis; however, the statistical results
of the latter concerning the subgrouping properties are robust
relative the error-boxes of data, unless some data will be 
modified drastically. 
Note the following two points.
On one hand, the relative velocity of
M31 subsystem with respect to the MW subsystem when
projected on the line joining MW to M31 is 
$-165 \pm 66$, confirming the conventional views.
On the other hand,
the 3D inner motions inside the Local Group, if significant,
are found to be surprisingly large
(M31 subsystem has a relative 3D velocity of amplitude
399 km s$^{-1}$ with respect to the MW subsystem).
It is interesting that the conclusion on the existence
of transverse velocity of MW relative to M31 has been
concluded previously by Peebles  (1994) using its least action
method (Peebles 1989).
The point is discussed below in term of the
convergence of the various luminous dipoles
with the one observed in the Cosmic Microwave
Background Radiation.

The CMB temperature
dipole, if interpreted as the signature of our
motion with respect to the rest frame of this radiation
(CMB frame), gives for the Milky Way a peculiar velocity
${\rm V}_{{\rm MW} \rightarrow {\rm CMB}}$
of 552 km s$^{-1}$ pointing toward the galactic
coordinates $l=266$ and $b=29$ (see Smoot et al. (1991, 1992),
Kogut et al. (1993)).
This MW motion 
in the CMB frame is traditionnally
split into two components 
\begin{equation}
{\rm V}_{{\rm MW} \rightarrow {\rm CMB}}=
{\rm V}_{{\rm MW} \rightarrow {\rm LG}}+
{\rm V}_{{\rm LG} \rightarrow {\rm CMB}}
\end{equation}
where 
${\rm V}_{{\rm MW} \rightarrow {\rm LG}}$ is the velocity of
the Milky Way relative to Local Group rest frame, which
originates from the internal non-linear dynamics governing the Local
Group and 
${\rm V}_{{\rm LG} \rightarrow {\rm CMB}}$ is the peculiar
motion of the Local Group as a whole in the CMB frame,
created by 
large scale mass fluctuations present in the Universe.

The latter can be infered in some way from the various luminous
dipoles found in the literature
(X-ray galaxies dipole by Miyaji \& Boldt (1990): $l=313$ and $b=38$;
Optical galaxies within 8000 km s$^{-1}$ dipole by Hudson (1993):  
$l=242$ and $b=49$ or
$l=231$ and $b=40$;
IRAS dipole, the least for shell  (142.8-157.3 Mpc) by
Plionis, Coles and Catelan (1993): 
$l=260.7$ and $b=39.1$;
Abell/AC0 clusters within 20 000 km s$^{-1}$ by Branchini\&Plionis (1996): 
$l=265$ and $b=16$ if corrected from Virgocentric flow).
Though there is no absolute convergence, it is noticed
that convergence level is improved if deeper samples of
extragalactic objects are involved. This fact indicates at least
that X-ray, optical and IR observed objects have approximatively
the same spatial distribution.

Conversion of luminous dipole in terms of the motion
${\rm V}_{{\rm LG} \rightarrow {\rm CMB}}$ 
of the Local Group in the CMB frame assumes the
following hypotheses :
\\\indent - H1) The luminous objects trace the large scale
mass density field.
\\\indent - H2) The linear approximation holds
(in particular, peculiar velocity remains
parallel to acceleration troughout the evolution
of large scale structures).
\\\indent - H3) The sample of objects under consideration
is at rest in the CMB frame.
\\
In what extent assumptions H1 and H2 are satisfied is yet an open
question, while results obtained from peculiar velocity analysis 
seem to challenge assumption H3
(Bulk flow of the shell of galaxies 
(within 3500-6500 km s$^{-1}$) by Rubin et al. (1976):
$v_B$=950 km s$^{-1}$ toward $l=308$ and $b=25$ in the CMB frame;
Bulk flow of clusters of galaxies (up to 15 000 km s$^{-1}$)
by Lauer\&Postman (1994): 
$v_B$=700 km s$^{-1}$ toward $l=340$ and $b=50$ in the CMB frame,
recently revisited by Graham (1996):
$v_B$=738 km s$^{-1}$ toward $l=330$ and $b=45$).

On the other hand, the peculiar motion of the Milky Way
in the Local Group rest frame
${\rm V}_{{\rm MW} \rightarrow {\rm LG}}$
can be obtained from the analysis
of the dynamics of the Local Group and its substructures.
Substracting it to the observed 
${\rm V}_{{\rm MW} \rightarrow {\rm CMB}}$
thus gives a local estimate of
${\rm V}_{{\rm LG} \rightarrow {\rm CMB}}$,
which can be compared with its values
extracted from dipoles analysis, as mentioned above. 

Herein, a rough kinematical model is adopted, assigning 
to the Local Group the mean motion of its main substructures
(i.e. MW and M31 subsystems, equally weighted). 
It gives 
${\rm V}_{{\rm MW} \rightarrow {\rm LG}}=
(7 \pm 303,15 \pm 155 ,-177  \pm 144)$
or $178$ km s$^{-1}$ toward $l=65$ and $b=-85$.
Finally, our local estimate of the LG peculiar
velocity in the
CMB frame yields 
${\rm V}_{{\rm LG} \rightarrow {\rm CMB}}
=(-41 \pm 303,-497 \pm 155 ,445  \pm 144)$
or $668$ km s$^{-1}$ in amplitude
pointing toward $l=265$ and $b=42$.
This result can be directly compared
with the well-known estimation by Yahil et al. (1977),
based on the Solar system motion relative
to the LG centroid :
${\rm V}_{{\rm LG} \rightarrow {\rm CMB}}=622$
km s$^{-1}$
toward $l=277$ and $b=30$ or 
$(66,-535,311)$ in galactic cartesian coordinates.

Thus we have obtained the LG motion in CMB frame in an alternative way.
Therefore it is remarkable that this result is in good agreement
with the result of Yahil et al (1977) within 1 $\sigma$ level. 
The existence of some discrepancy between these two values, if significant,
can be interpreted as follows.
The LG centroid had been defined by specific choice of the
main and satellite populations and with further search of the best-fit
solution for the Solar system motion. 
In the present analysis we have found  statistically
significant indication of the bulk flow of the two main subsystems of LG,
which can influence the definition of its centroid, and hence, the final 
result. 

The understanding of the cause of each discrepancy is the main problem
to be solved. The LG substructure's 3D dynamics, as discussed above, 
could be essential for that problem.
The fact of the existence of the bulk flow of the substructures should be
crucial also while studing the past and future evolution of the Local Group.

Moreover, bulk flows can be common properties of subgroups of clusters
of galaxies (Gurzadyan\&Mazure, 1997) -- galaxy associations, thus
reflecting the role of merging and other basic trends in
the formation mechanisms of the clusters.

\subsection*{Acknowledgements} 
S. Rauzy recognizes the hospitality of the Centre de Physique
Th\'eorique.
V.G. Gurzadyan was supported by Royal Society and French-Armenian PICS.

\appendix
\section*{Appendix A : The bulk flow statistic}
\label{appendix}

In this appendix,
we use the maximum likelihood
technique in order to derive the statistics of the $3$ 
components of the bulk flow. Quantities, such as 
the variance of these estimators or
the velocity dispersion estimate, are obtained by using
numerical simulations.

We define the statistical model as follows.
It is assumed, that
the data sample consists of $N$ independent objects, which
follow the theoretical probability density $dP_{\rm th}$ of
Eq. (2) : 
$$
dP_{\rm th}=\phi(v_1,v_2,v_3)~dv_1\,dv_2\,dv_3 \times \rho(r,l,b)~
r^2\cos b \,dr\,dl \,db
$$
where 
$\phi(v_1,v_2,v_3)$ is the 3D velocity distribution function (in
galactic cartesian coordinates) and $\rho(r,l,b)$ is the 3D spatial
distribution function (in galactic coordinates).
Hereafter, we assume that
the 3D velocity distribution function of
the sample adopts the following form 
$$
\phi(v_1,v_2,v_3)=
g(v_1;v_x,\sigma_v)\times
g(v_2;v_y,\sigma_v)\times
g(v_3;v_z,\sigma_v)
$$
where $g(x;x_0,\sigma_x)$ is a Gaussian of mean $x_0$ and of dispersion
$\sigma_x$. The velocity field  of the sample is thus
split into a mean 3D velocity
${\bf v}_B=(v_x,v_y,v_z)$ (i.e. a bulk flow) plus a 3D random
component, isotropic Maxwellian, centered on ${\bf 0}$ and
of velocity dispersion $\sigma_v$. 

For a given object, the observables are the galactic longitude $l$
and latitude $b$ and the radial velocity $v_r$ given
in the galactocentric frame (MWR frame) :
\begin{equation}
\label{radialvelocity}
v_r = {\bf v}.{\bf {\hat r}} =
v_1 \cos l \cos b + v_2 \sin l \cos b + v_3 \sin b 
\end{equation}
where ${\bf {\hat r}}=
(\cos l \cos b,\sin l \cos b,\sin b)=
({\hat r}_1,{\hat r}_2,{\hat r}_3)$ is the line-of-sight direction
and ${\bf v}=(v_1,v_2,v_3)$ the 3D velocity of the galaxy
in the MWR frame. 
By successively integrating
over the distance $r$ and over 2 components of
the 3D velocity (say, of $v_1$ and $v_2$), we
express the observed probability
density $dP_{\rm obs}$ in terms of the observables :
$$
dP_{\rm obs} =
g(v_r;v_x \cos l \cos b + v_y \sin l \cos b 
+ v_z \sin b,\sigma_v)~dv_r 
$$
\begin{equation}
\label{dpobs1}
~~~~~~~~~~~\times \eta(l,b)~\cos b\,dl\,db
\end{equation}
where $\eta(l,b)$ is 
the projected angular distribution function
of the objects under consideration 
(i.e. $\eta(l,b) = \int \rho(r,l,b)~r^2\,dr$).
Note, that the information on the distances $r$ of
the galaxies is not used.
For a data sample of N galaxies with
measured $\{l^k,b^k,v_r^k\}_{k=1,N}$,
the efficient part of the natural logarithm of the
likelihood function
${\cal L}= 
{\cal L}(v_x,v_y,v_z,\sigma_v)$
reads thus :
\begin{equation}
\label{lf}
{\cal L}=
-\ln \sigma_v 
-\frac{1}{N}\sum_{k=1}^{N}
{{(v_r^k-v_x
{\hat r}_1^k 
 - v_y
{\hat r}_2^k 
- v_z 
{\hat r}_3^k 
)}\over{2 \sigma_v^2}}^2
\end{equation}
where ${\hat r}_1^k = \cos l^k \cos b^k$,
$ {\hat r}_2^k = \sin l^k \cos b^k$ and
$ {\hat r}_3^k = \sin b^k$ (see Eq. (\ref{radialvelocity})).
Maximizing ${\cal L}$ with respect to $v_x$, $v_y$ and $v_z$
gives the following set of equations 
\begin{equation}
\begin{tabular}{l}
$
\langle
{\hat r}_1 (v_r-v_x {\hat r}_1 - v_y {\hat r}_2 
- v_z {\hat r}_3)
\rangle
=0
$
 \\
 \\
$
\langle
{\hat r}_2 (v_r-v_x {\hat r}_1 - v_y {\hat r}_2 
- v_z {\hat r}_3)
\rangle
=0
$
 \\
 \\
$
\langle
{\hat r}_3 (v_r-v_x {\hat r}_1 - v_y {\hat r}_2 
- v_z {\hat r}_3)
\rangle
=0
$
\end{tabular}
\end{equation}
where $\langle . \rangle$ denotes the average on the sample
(for example $\langle v_r {\hat r}_2 \rangle= 
1/N \sum^{N}_{k=1} v_r^k {\hat r}_2^k =
1/N \sum^{N}_{k=1} v_r^k \sin l^k \cos b^k$).
This set of linear equations can be rewritten 
\begin{equation}
{\bf M} 
\left[ 
\begin{tabular}{r}
$ v_x $\\
$ v_y $\\ 
$ v_z $ 
\end{tabular} 
\right]
~=~
\left[ 
\begin{tabular}{r}
$\langle v_r {\hat r}_1 \rangle$ \\
$\langle v_r {\hat r}_2 \rangle$\\ 
$\langle v_r {\hat r}_3 \rangle$ 
\end{tabular} 
\right]
\end{equation}
where ${\bf M}$ is the 3$\times$3 symmetric matrix :
\begin{equation}
\label{matrixM}
{\bf M}
~=~
\left[ 
\begin{tabular}{rrr}
$\langle {\hat r}_1 {\hat r}_1  \rangle$ &
$\langle {\hat r}_1 {\hat r}_2  \rangle$ &
$\langle {\hat r}_1 {\hat r}_3  \rangle$\\ 
$\langle {\hat r}_1 {\hat r}_2  \rangle$ &
$\langle {\hat r}_2 {\hat r}_2  \rangle$ &
$\langle {\hat r}_2 {\hat r}_3  \rangle$\\ 
$\langle {\hat r}_1 {\hat r}_3  \rangle$ &
$\langle {\hat r}_2 {\hat r}_3  \rangle$ &
$\langle {\hat r}_3 {\hat r}_3  \rangle$ 
\end{tabular} 
\right]
\end{equation}
Finally, the unbiased statistics ${\bf {\tilde v}}_B=({\tilde v}_x,
{\tilde v}_y,{\tilde v}_z)$ of the 3D bulk flow 
${\bf v}_B=(v_x,v_y,v_z)$ is obtained by inverting
the 3$\times$3 matrix ${\bf M}$ :
\begin{equation}
\left[ 
\begin{tabular}{r}
$ {\tilde v_x} $\\
$ {\tilde v_y} $\\ 
$ {\tilde v_z} $ 
\end{tabular} 
\right]
~=~
{\bf M^{\small -1}} 
\left[ 
\begin{tabular}{r}
$\langle v_r {\hat r}_1 \rangle$ \\
$\langle v_r {\hat r}_2 \rangle$\\ 
$\langle v_r {\hat r}_3 \rangle$ 
\end{tabular} 
\right]
\end{equation}
This bulk flow statistic is robust.
It does not depend on the distance of galaxies
nor on the projected angular
distribution function $\eta(l,b)$ of the sample.
However, the accuracy of the
${\tilde v_x}$,
${\tilde v_y}$ and 
${\tilde v_z}$ estimators depends on $\eta(l,b)$
and on the velocity dispersion $\sigma_v$ of the dynamical
system. We thus, compute the variance of these estimators
by using numerical simulations which are supposed to mimic
the real data.

The derivation of the velocity dispersion estimator
${\tilde \sigma}_v$ is not straightforward. Maximizing
the efficient part of the likelihood function $\cal L$
of Eq. (\ref{lf}) with respect to $\sigma_v$, we have the following
equation 
\begin{equation}
\label{velocitydispersion}
\sigma_v^2=
\langle
(v_r-v_x {\hat r}_1 - v_y {\hat r}_2 
- v_z {\hat r}_3)^2
\rangle
\end{equation}
Unfortunatly, the velocity dispersion estimator $\sigma_v^\star$ :
\begin{equation}
\label{sigmavstar}
\sigma_v^\star\,^2=
\langle
(v_r-{\tilde v}_x {\hat r}_1 - {\tilde v}_y {\hat r}_2 
- {\tilde v}_z {\hat r}_3)^2
\rangle
\end{equation}
obtained while replacing the bulk flow ${\bf v}_B$ by its
estimate ${\bf {\tilde v}}_B$, is biased. The reasons
of this bias are twofold.
On one hand, the variance of the 
${\tilde v_x}$,
${\tilde v_y}$ and 
${\tilde v_z}$ estimates leads to enhance the value
of $\sigma_v^\star$ and thus, to overestimate the velocity
dispersion $\sigma_v$.
On the other hand, because of the finite size of the sample,
the random variables  
${\tilde v_x}$,
${\tilde v_y}$, 
${\tilde v_z}$ and the radial velocity
$v_r$ of each objects are correlated. This
feature contributes to underestimate
$\sigma_v$ when using the 
$\sigma_v^\star$  statistic. 
As a matter of fact,
the unbiased estimator of the velocity dispersion 
depends on the accuracy of the bulk flow estimate 
and thus on the velocity dispersion itself.
In this paper, we have applied a computational iterative
process on numerical simulations which furnishes
an unbiased value for $\sigma_v$.

\bsp

\end{document}